\documentclass{elsart}

\newcommand{\W}{14cm}

\usepackage{natbib}

\usepackage{graphicx,amssymb}

\journal{geothermics}

\begin{document}

\begin{frontmatter}

\title{Flow channelling in a single fracture induced by shear displacement}
\author{H. Auradou$^1$}
\author{G. Drazer$^{2,3}$}
\author{A. Boschan$^{1,4}$}
\author{J.P. Hulin$^1$}
\author{J. Koplik$^2$}
\address{$^1$Laboratoire Fluide, Automatique et Syst{\`e}mes 
Thermiques, UMR 7608, Universit{\'e} Paris 6 and 11, 
B{\^a}timent 502, Campus Paris Sud Paris Sud, 91405 Orsay 
Cedex, France\\$^2$Benjamin Levich Institute and Department of Physics,
City College of the City University of New York, New York, NY
10031.\\$^3$ Dept. Chemical \& Biomolecular Engineering, The Whiting
School of Engineering Johns Hopkins University, 218 Maryland Hall, Baltimore, MD 21218-2681\\$^4$Grupo de Medios Porosos, Facultad de Ingenieria, Paseo Colon 850, 1063 Buenos-Aires, Argentina.}

\begin{abstract}
The effect on the transport properties of fractures of a relative shear displacement  $\vec u$ of rough walls with complementary self-affine surfaces has been studied experimentally and numerically. 
The shear displacement $\vec u$ induces  an anisotropy of the aperture field with a correlation length scaling as $u$ and significantly larger in the direction perpendicular to $\vec u$. This reflects the appearance of long range channels perpendicular to $\vec u$ resulting in a higher effective permeability for flow in the direction perpendicular to the shear.
Miscible displacements fronts in such fractures are observed experimentally to display a self affine geometry of characteristic exponent directly related to that of the rough wall surfaces. A simple  model based on the channelization of the aperture field allows to reproduces  the front geometry when the mean flow is parallel to the  channels created by the shear displacement. 
\end{abstract}
\end{frontmatter}

\section{Introduction}
The geothermal reservoir of Soultz-sous-For\^ets, like most geological systems, contains fracture networks 
providing preferential paths for fluid flow and identifying these paths is crucial in 
geothermal applications. More precisely, the number of hydraulically connected fractures  and their spatial distribution are key factors in determining the rate at which hot water can be produced or fluids can be re-injected into the system \citep{Genter97,Dezayes04}. 
Transport in these systems can be stimulated by the massive injection of water at high flow rates which is commonly used to increase the overall transmissivity : both borehole and microseismic observations have demonstrated that the relative shearing of the fracture walls is a key factor in this enhancement  \citep{Genter97,Evans05}.  Another important issue is the influence of the shear on the transport of suspended particles and dissolved species.

The present work analyzes the influence of shear displacements of the fracture walls on the pore space structure and, therefore, on the transport properties. The case of fractures with a self-affine geometry is specifically investigated : this  statistical model describes indeed successfully a broad variety of fractured materials (see \cite{Bouchaud03} for a review). This model which will be described more precisely in section~\ref{sec:correlation} accounts also well for the long range correlations in the topography of such surfaces.

An important factor in these processes is the anisotropy of the permeability field induced by the shear displacement of the fracture walls : the permeability is highest for  flow perpendicular to the shear displacement vector ${\vec u}$ and lowest  along it~\citep{Gentier97,Yeo98}. 
This anisotropy has been suggested by~\cite{Auradou05} to be associated to the appearance, in the aperture field, of ridges perpendicular to  ${\vec u}$. These ridges partly inhibit flow parallel to ${\vec u}$ while they create long preferential flow channels connected in series in the perpendicular direction :  this leads to model the pore space as long parallel channels extending over the length of the fracture. In the present  work, the validity of this simple description of the void space will be confirmed by  a detailed study of the geostatistical characteristics of the aperture field. 
It will be shown that both 
the anisotropy of the aperture field and the presence of large scale surface structures spanning the whole surface are needed to account for the anisotropy. 

These analysis  will then be extended to  the spreading of a tracer advected by the flow. This process  is largely determined by the structure of the flow field which would, at first, seem to be very complex. 
However, both experimental studies \citep{Auradou01} and numerical simulations \citep{Drazer04} suggest that, for shear displacements $\vec u$ small compared to the fracture size and for flow perpendicular to $\vec u$, the flow lines are mostly parallel to the mean flow  
the velocity field remains correlated over distances of the order of the fracture length.  
Finally, the correlation of the velocities of  tracer  particles advected in the fracture will be shown to have also self-affine characteristics, and a model accounting for the geometry of  miscible displacement fronts in such fractures will be presented. 

\section{Experimental setup and numerical models} 
\begin{figure}
\includegraphics*[width=\W]{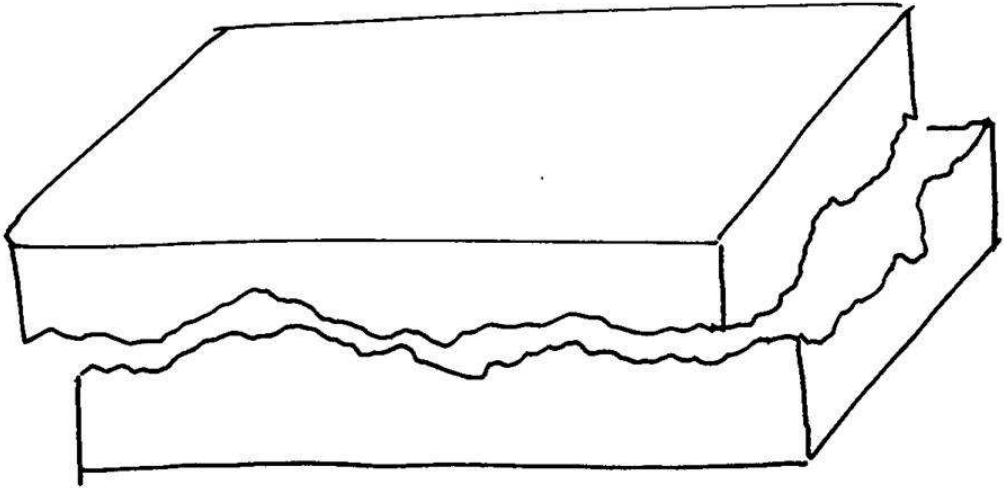}
\caption{Schematic view of the experimental fracture models. The fracture walls are complementary surfaces with the same roughness geometry and self-affine exponent $\zeta\sim0.8$ as natural granite fractures. The model walls are shifted vertically to determine the mean aperture and horizontally to model relative shear displacements.}
\label{fig:fig1}
\end{figure}

Both in the experiments and the numerical simulations, the fractures are modeled as the gap between two  complementary rough surfaces, as shown in Fig. \ref{fig:fig1}. 
Experimentally, these surfaces are transparent casts of granite or fracture surfaces that retain their {\it self affine} geometry.
 The experimental setup allows us to 
control both the normal distance and the lateral displacement between 
the two facing wall surfaces (a detailed description has been given by \citet{Auradou01,Auradou05} ).
The normal distances investigated range from $500 \mu m$ to $1\ mm$, i.e. of the same order 
of magnitude as the mean aperture of fractures in the Soultz Geothermal reservoir \citep{Sausse02}. 
The lateral displacement mimics a relative shear displacement of the walls and induces spatial 
variations of the aperture. The magnitude of the shear displacements ranges from zero (no shear configuration) to a few millimeters when the lateral displacement brings the opposite surfaces in contact. The   shear displacement amplitudes are of the same order of magnitude than 
those estimated by \citet{Evans05} after an hydraulic stimulation. The resulting 
aperture fluctuations  are also similar to those reported by Sausse \citep{Sausse02} 
for fractures in the Soultz reservoir.   
\begin{figure}
\includegraphics*[width=\W]{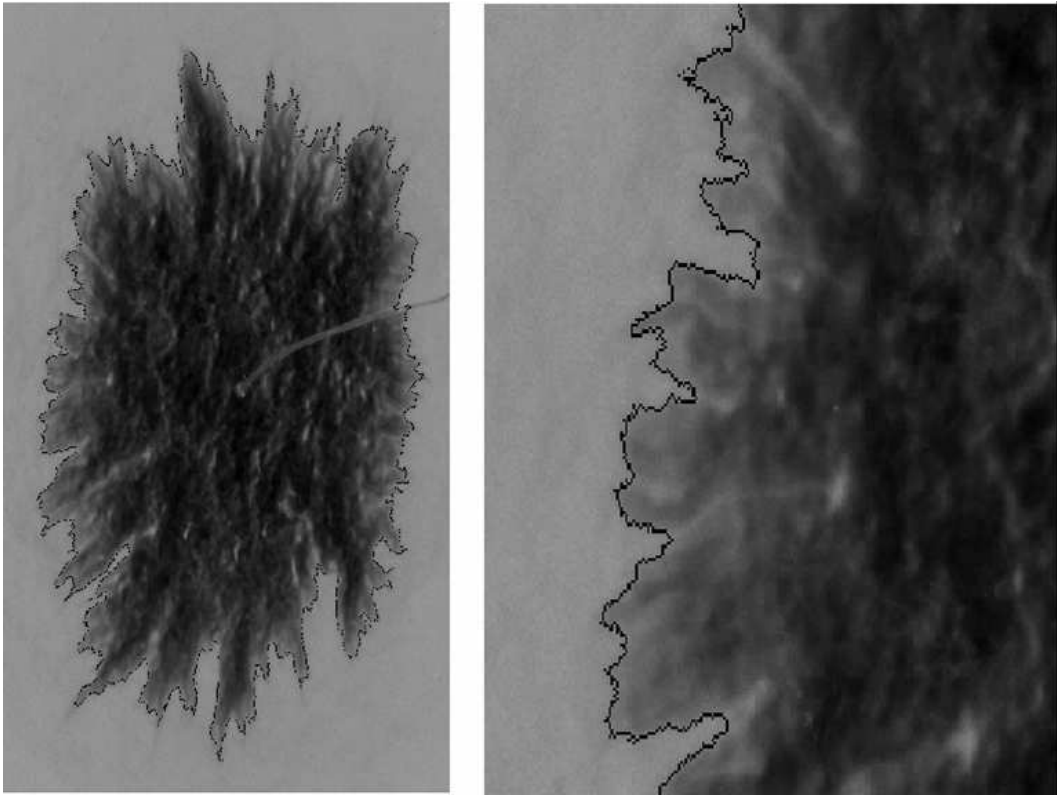}
\caption{a) Experimental pictures of the radial injection of a dyed fluid into a fracture saturated by a miscible, transparent one (picture size : $20\ cm \times 20\ cm$. b) Detail of the invasion front.}
\label{fig:fig2}
\end{figure}

The fracture model is fitted inside a parallelepipedic basin : both are initially
 carefully saturated with a transparent solution so that no air bubble gets trapped inside the setup. The same solution, but dyed, is then injected through a small borehole in the middle of the 
upper epoxy cast and pictures of the dye invasion front are obtained at regular time intervals. 
Figure \ref{fig:fig2} displays a typical invasion front geometry : its larger elongation in the direction perpendicular to the shear reflects the influence of the permeability anisotropy discussed above.
Moreover, as discussed by \citet{Auradou05},  the ratio of the elongations of the front parallel 
and perpendicular to the shear is directly related to the ratio of the permeabilities along these two
same directions.

\begin{figure}
\includegraphics*[width=\W]{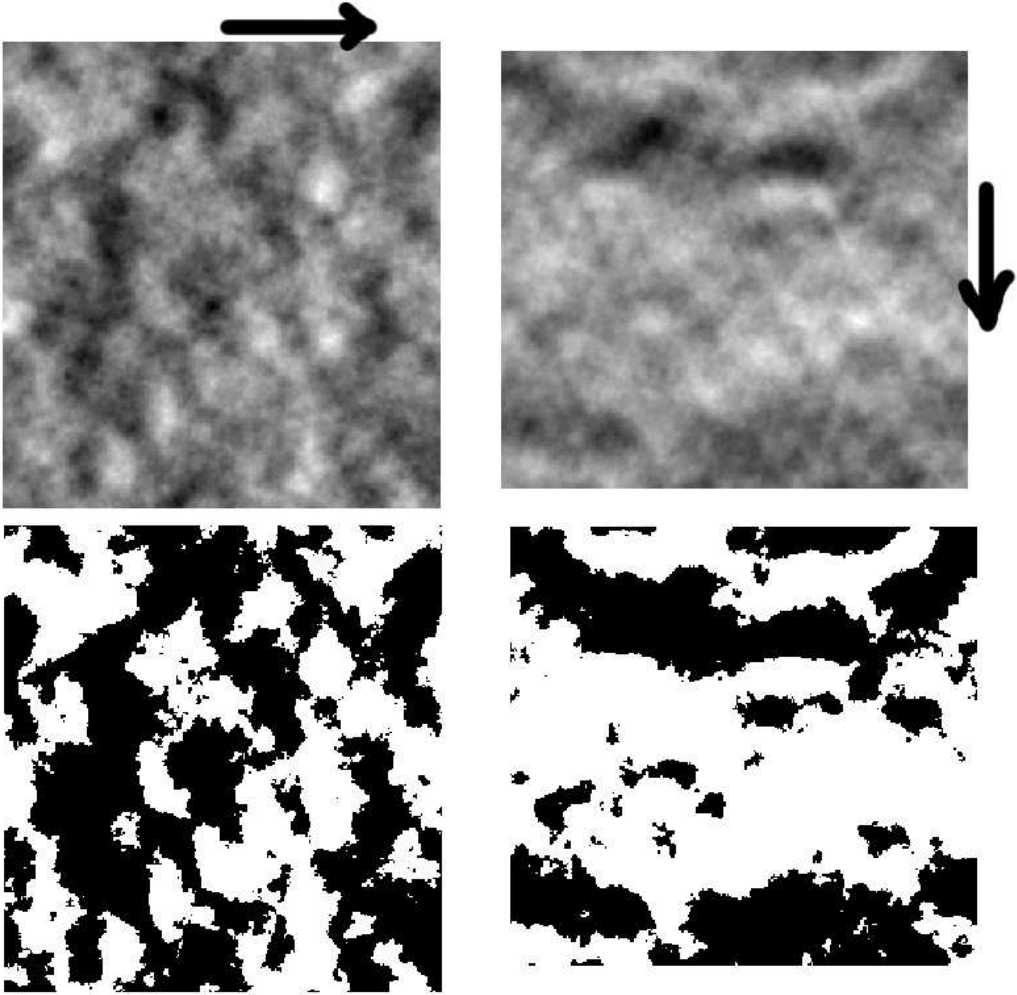}
\caption{Numerical aperture fields ($51.2\ mm \times 51.2\ mm$) obtained for a same magnitude of the shear displacement  ${\vec u}$ ($u = 0.8\ mm$) along two different perpendicular directions (arrows). Upper images - aperture fields coded in  grey levels. Lower images : binarized images corresponding to a $1\,mm$ threshold equal to the mean aperture (lower apertures appear in white).
Left : $S=0.16$, $K_\perp/K_0=1.05$ and $K_\parallel/K_0=0.822$ ($S$ = ratio of the variance of the aperture fluctuations by the mean aperture, $K_\parallel$ and $K_\perp$ : permeability values for flow parallel and perpendicular  to the shear displacement ${\vec u}$, $K_0$ = permeability for $u = 0$.).
Right : $S=0.21$, $K_\perp/K_0=1.00$ and $K_\parallel/K_0=0.930$.}
\label{fig:fig3}
\end{figure}

As mentioned above, aperture variations due to the lateral displacement of the fracture walls strongly perturb the flow field. The numerical simulation of  the above experiments 
requires therefore an accurate 3D description of the fracture geometry and of the velocity field : 
the lattice Boltzmann technique has been selected for that purpose because it is well adapted to complex geometries \citep{Drazer00,Drazer02}. 
A large cubic lattice ($1024  \times  1024 \times 20$ sites) is used to reproduce a model fracture 
of mean aperture $1\ mm$ and area $51.2 \times 51.2\ mm$ (one fourth of the experimental area).  
Using these numerical simulations, the variability of the results from one fracture 
to another has been investigated and the permeabilities for flow both parallel 
and perpendicular to the shear direction have been determined (note that only their ratio can be measured experimentally).

Figure \ref{fig:fig3} compares aperture fields obtained in numerical simulations and corresponding to relative displacements of the same complementary surfaces in two directions at right angle. Permeability values  computed in these two cases for constant pressure gradients  parallel and perpendicular to the shear are listed in the caption.  One observes the appearance of near continuous streaks perpendicular to the shear displacement and marking preferential flow paths. 
The appearance of these paths is reflected in the difference between the permeabilities $K_\perp$ and $K_\parallel$ for flow respectively parallel and perpendicular to them.
More precisely,  the permeability $K_\perp$ for flow perpendicular to the displacement ${\vec u}$ increases with $u$  while $K_\parallel$ decreases. These variations are observed~\citep{Auradou05} to be related to the permeability $K_0$ for $u = 0$ and to the ratio $S$ of the  variance of the aperture fluctuations, $\sigma_a$ by  the mean aperture, $a_0$ through the approximate equations  :
\begin{eqnarray}
\frac{K_\perp}{K_0} &=& 1 + 3 A S^2, \nonumber \\
\frac{K_\parallel}{K_0} &=& 1 - 6 B S^2 + O(S^4),
\label{eq:eq1}
\end{eqnarray}
 The two constants, $A$ and $B$, 
depend on the specific statistical properties of each individual fracture surfaces but satisfy the relations:  $A<1$ and $B<1$ .

As stated above, the binarized aperture fields shown in Fig. \ref{fig:fig3} displays ridges perpendicular to the dispacement  ${\vec u}$ and extending over a large part of the fracture length parallel to the flow. This observation has been used as the basic ingredient of 
a simple model in which the aperture field is replaced by a set of parallel channels, each of constant aperture and perpendicular to the shear. This model captures the linear dependence of the permeabilities 
with respect to $S^2$ and gives a good estimate of the respective upper and lower bounds of $K_\perp$ and $K_\parallel$  \citep{Zimmerman91,Zimmerman96,Auradou05}. 
These two limits correspond to $A=1$ and $B=1$ in Eq. (\ref{eq:eq1}).
This simple model takes advantage of the fact that the correlation lengths in the aperture field are highly anisotropic and much smaller parallel to ${\vec u}$ than perpendicular to it.  

 In addition to its global anisotropy, the experimental pictures displayed in Figure \ref{fig:fig2} show that the  front  between the transparent and dyed fluids is locally quite tortuous. More precisely,  its geometry has been found experimentally to be self-affine with a characteristic exponent close to that of the roughness of the fracture surfaces~\citep{Auradou01}. 
This result has been verified numerically by \citet{Drazer04} for two values  $\zeta = 0.8$ and $0.5$ of the characteristic  exponent $\zeta$~\citep{feder88} of the fracture wall surfaces (note that $\zeta$ has the same "universal" value $0.8$ for the  basalt and granite surfaces modeled in the present work and its influence could not therefore  be investigated experimentally). 
 
The spatial correlation of the aperture and velocity fields in these fractures will now be analyzed more quantitatively using standard geostatistical tools. A model based on these caracteristics will then be discussed in section \ref{sec:front} and shown to capture the key features of miscible displacement processes studied experimentally and numerically. 

\section{Spatial correlation of the aperture field}
\label{sec:correlation}
The following interpretations are realized on 2D maps with $770\times760$ pixels measured experimentally  on the epoxy surfaces  by a mechanical profilometer; the spacing between pixels  is  $250\,\mu m$ in both directions of the mean plane of the fracture and the vertical resolution on the local height $h(x,y)$ is $10\, \mu m$. 
The statistical properties of the surface
roughness have determined from these maps and, as stated above, can be characterized in terms of a  {\it self-affine} geometry.  The main property of a self-affine surface is that it remains statistically unchanged under the transformation\,: 
\begin{eqnarray}
x &\rightarrow& \lambda x, \nonumber \\
y &\rightarrow& \lambda y, \nonumber\\
h(x,y) &\rightarrow& \lambda^\zeta h(x,y). \nonumber
\end{eqnarray}
Where $(x,y)$ are the coordinates in the mean surface plane, $h(x,y)$ is the local height of the surface and $\zeta$
is refered to as the Hurst, roughness, or self-affine exponent \citep{feder88}. For the experimental model fractures, one has  $\zeta = 0.75 \pm 0.05$ \citep{Auradou05}, in agreement with published values for granite surfaces~\citep{Bouchaud03}.
In the simulations, the {\it self-affine} surfaces are generated numerically (see \citet{Drazer02} for a detailed description of the method). 

In the present work, we consider fractures made of complementary walls that are separated both vertically in order to 
open the fracture and laterally in order to mimic shear displacements (if any). Under such 
conditions, if $a_0$ is the mean aperture and $\vec{u}$ the relative shear displacement  in the plane of the fracture, the  local aperture at a point $\vec{r}$ satisfies :
\begin{equation}
a(\vec{r}) = h(\vec{r}) - h(\vec{r}+\vec{u}) + a_0.
\label{eq:a}
\end{equation}
This equation allows to determine the aperture field $a({\vec r})$ for a given shift ${\vec u}$ from the experimental surface profile map $h({\vec r})$.
The correlations in the aperture field are characterized by the
 following correlation function, also called semivariance \citep{Kitanidis97}:
\begin{equation}
  \gamma(\vec{\delta}) = \langle (a(\vec{r}) - a(\vec{r}+\vec{\delta}))^2\rangle,
\label{eq:delta}
\end{equation}
which measures the spatial correlation of the aperture field between two points separated by the 2D vector $\vec{\delta}$. 

Combining Eqs.~\ref{eq:a} and \ref{eq:delta}, one sees that the value of $\gamma$ is determined 
by the correlation between the heights of the surface at four  points  located at the corners of a parallelogram of 
sides $\vec{u}$ and $\vec{\delta}$. The orientations of the lag $\vec \delta$ respectively parallel and perpendicular to the shear displacement ${\vec u}$ are of special interest (in the following, they are respectively referred to as $\parallel$ and $\perp$). 
\begin{figure}
\includegraphics*[width=\W]{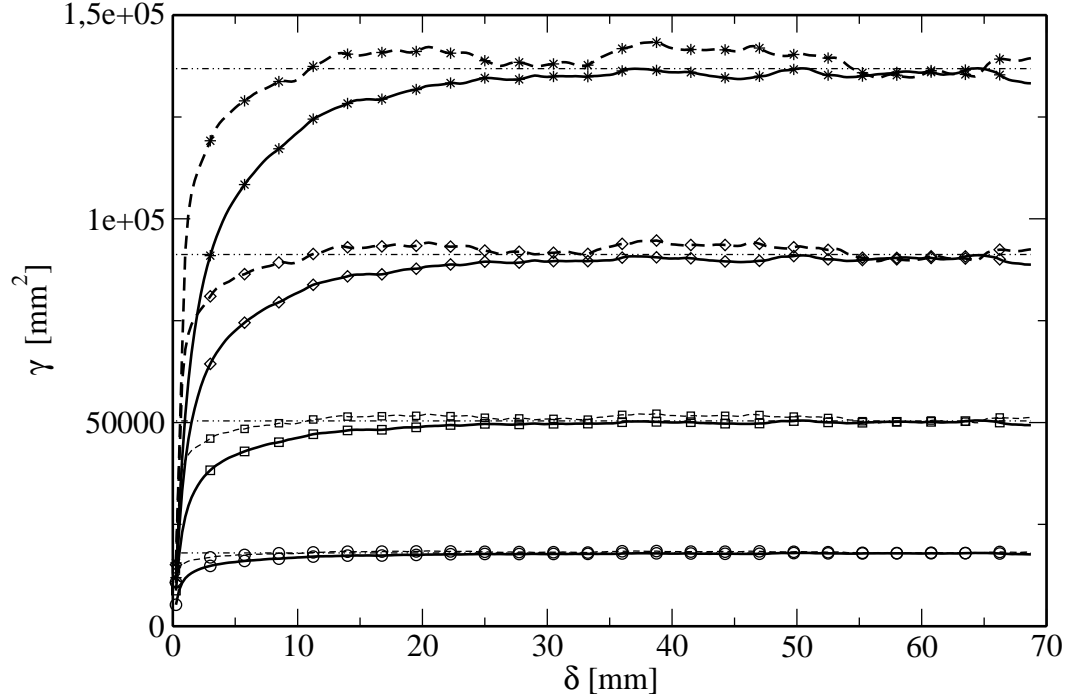}
\caption{Semivariograms of aperture fields corresponding to shift amplitudes $u$ equal to : ($ 0.25\ mm$, $\circ$), ($0.5\ mm$,$\square$), 
($0.75\ mm$,$\Diamond$), and ($1\ mm$, $\star$). Dotted lines : $\gamma_\parallel$ (correlation 
along the direction of ${\vec u}$). Solid lines : $\gamma_\perp$ (correlations in the direction perpendicular to ${\vec u}$). 
Horizontal dashed lines correspond to twice the aperture variance for the same $u$ values.}
\label{fig:semivariograms}
\end{figure}
\begin{figure}
\includegraphics*[width=\W]{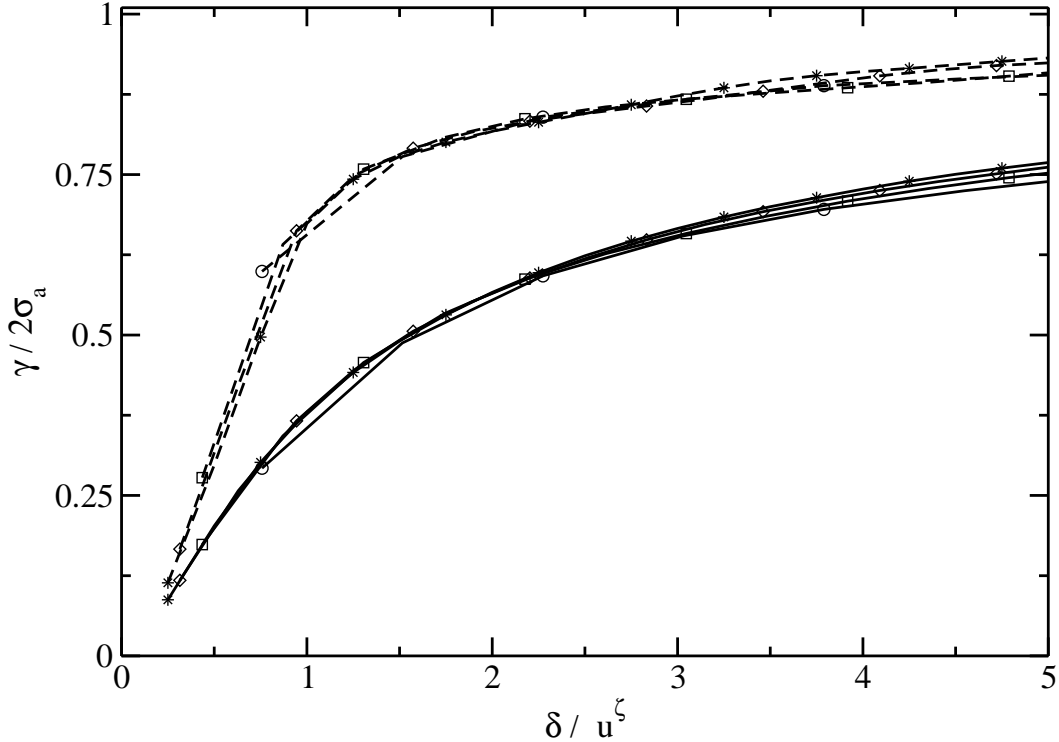}
\caption{Semivariograms of aperture fields normalized by $A u^{2\zeta}$ ($\zeta=0.75$ and $A\,=\,137000\,mm^{2(1-\zeta}$ as function of the normalized lag $\delta / u$. Shear amplitude  $u$ values  : ($0.25\ mm$, $\circ$), 
($0.5\ mm$,$\square$), ($0.75\ mm$,$\Diamond$), and ($1\ mm$, $\star$). Dotted lines : $\gamma_\parallel$ (correlation 
along the direction of ${\vec u}$). Solid lines : $\gamma_\perp$ (correlations in the direction perpendicular to ${\vec u}$). Note that the horizontal range is narrower than in 
Fig.~\ref{fig:semivariograms} so that the overshoot effect is not visible.}
\label{fig:correlation}
\end{figure}

Figure \ref{fig:semivariograms} displays variations with $\delta$ of  the semivariances $\gamma_\perp$ and $\gamma_\parallel$ for aperture fields computed from a same experimental surface profile map $h({\vec r})$ but for different shear displacements ${\vec u}$. For
small values of the lag $\vec{\delta}$, the apertures at $\vec{r}$ and $\vec{r}+\vec{\delta}$ are very similar (high
spatial correlation at short distances) and the value of $\gamma$ is low. As $\delta$
increases, so does $\gamma$ because the correlation decreases. 
As $\delta$ increases further to a value higher  than the correlation length of the aperture field, the apertures at $\vec{r}$ and $\vec{r}+\vec{\delta}$
become uncorrelated; then, from Eq. (\ref{eq:delta}), $\gamma_\perp$ and $\gamma_\parallel$ should reach a same constant saturation value equal to  twice the aperture variance $\sigma_a=\langle (a(x,y) - a_0)^2\rangle$. The values of $\sigma_a$ corresponding to the different shear amplitudes $u$ are plotted as horizontal dashed lines on Fig.~ \ref{fig:semivariograms} :  they represent indeed the limit of both $\gamma_\perp$ and $\gamma_\parallel$ at high $\delta$ values. Also, as  previously shown 
by \cite{Auradou05}, $\sigma_a$ increases with the magnitude $u$ of the shear displacement : in the range of $u$ values used in Fig.\,\ref{fig:semivariograms}, one verifies experimentally that $\sigma_a = A u^{2\zeta}$ with a best fit corresponding to $\zeta=0.75$ and $A\,=\,137000\,mm^{2(1-\zeta)}$.

Even though the limits of $\gamma_\perp$ and $\gamma_\parallel$ are the same at large $\delta$ values  for a given shear displacement $u$, the way in which this saturation is reached differs very much for the two semivariances due to the anisotropy of  the aperture field.
The semivariance $\gamma_\perp$ for $\vec{\delta}$ perpendicular to  $\vec{u}$ never exceeds its saturation value and reaches it  in an {\it overdamped} way whereas $\gamma_\parallel$ ($\vec{\delta}$ parallel to  $\vec{u}$)  displays an overshoot  reflecting  a local anticorrelation.
These different behaviors result directly from the large scale anisotropic structures observed above on Fig. \ref{fig:fig3}. Let us consider  the aperture at two points separated by a small distance $\vec{\delta}$ perpendicular to the shear displacement (and therefore parallel to the length of the structures) : they will most likely be both either above or below the average aperture if they are inside a same large scale structure. 
Then, the corresponding value of aperture semivariance will be less than 
$\sigma_a$. On the other hand, two points with a separation $\vec{\delta}$ parallel to $\vec u$ have a high probability to correspond to different structures of  the binarized picture : some amount of anti correlation can therefore be expected.  

Let us now investigate the parameters determining the variation of $\gamma$ in the transition regime. For that purpose, we have seeked to collapse  the different curves of Fig.\ref{fig:semivariograms} onto master curves. In Fig.~\ref{fig:correlation}, $\gamma_\perp$ and $\gamma_\parallel$ are normalized by $A u^{2\zeta}$ which represents their limit at large values of $\delta$ while,  
for the horizontal scale,  $\delta/u$ is a natural reduced coordinate (see \citet{Plouraboue95}). These two normalizations allow indeed to collapse in Fig. \ref{fig:correlation} the different variations  into  two independent master curves, corresponding to the two orientations of $\vec \delta$ with respect to
$\vec u$. This demonstrates, that, in all cases,  the characteristic distance over which the aperture field gets decorrelated is proportional to the shift $u$ : in addition, the ratio of this distance by $u$ is significantly larger for $\delta$ perpendicular to the shear than parallel to it. While this latter result will remain valid, the actual values of both distances may vary significantly from one sample to another.

Fig. \ref{fig:correlation} also shows that, 
for small lag values $\delta/u \ll 1 $), the semivariograms increase linearly with $\delta/u$ ($\gamma/u^\zeta \propto \delta/u$). This indicates that the random field is not as smooth as in the case of a  Gaussian covariance function for which $\gamma \propto \delta^2$.

\begin{figure}
\includegraphics*[width=\W]{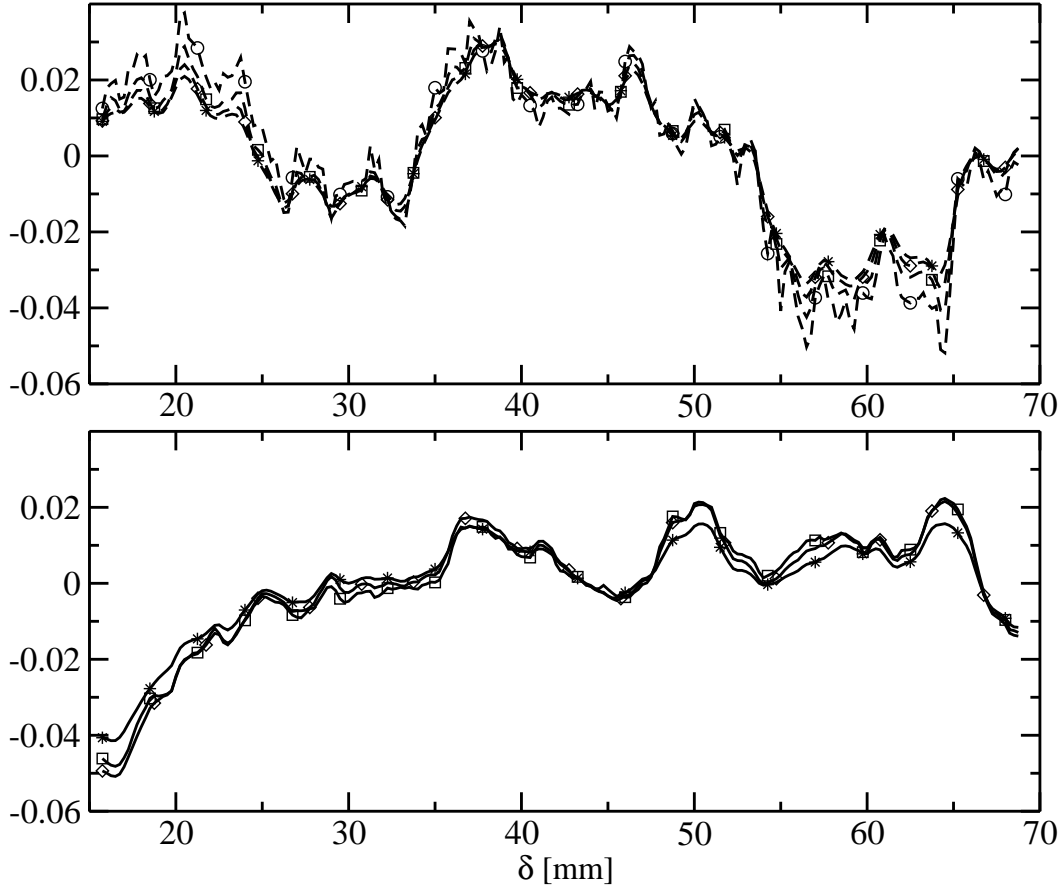}
\caption{Variations of  $(\gamma_{\parallel} / \bar{\gamma}_{\parallel} -1)$ (top) and $(\gamma_{\perp} / \bar{\gamma}_{\perp} -1)$ (bottom)
normalized by $u^\zeta$ ($\zeta=0.75$) as a function of $\delta$. Shear amplitudes $u$ : ($0.25\ mm$, $\circ$), ($0.5\ mm$,$\square$), ($0.75 mm$,$\Diamond$),  and ($1\ mm$, $\star$).}
\label{fig:normal}
\end{figure}
At large distances   $\delta \gg u$, the semivariance $\gamma$ still displays large fluctuations with excursions above and below the  asymptotic value $\sigma_a$ : these are visible  in Fig. \ref{fig:normal} that represents the normalized deviations $(\gamma - \bar{\gamma})/\bar{\gamma}$ of $\gamma$ from its average value $\bar{\gamma}$ over all distances $\delta > 15\,mm$.  It is observed that the fluctuations coincide provided $\delta$ is used as the horizontal scale (and not  the normalized lag $\delta/u$ as in Fig.~\ref{fig:correlation}). This suggests that these variations reflect the  large scale underlying structures displayed in Fig. \ref{fig:fig3} and that their location is independent on the magnitude $u$ of the shear displacement (these fluctuations will however be very variable from a sample to an other one).  The variation of the amplitude of  the fluctuations with $u$ is however non trivial : the normalized deviations $(\gamma - \bar{\gamma})/\bar{\gamma}$ had to be divided by  $u^\zeta$ to obtain the collapse of  Fig. \ref{fig:normal}. This indicates that there is some influence of the local self-affine structure on these fluctuations; if they were only due to the large scale channels, the normalized deviations would be proportional to $u$.
\section{Correlation between fracture aperture fields and fluid displacement fronts}
\label{sec:front}
This section is devoted to the geometry of the front between two miscible fluids (or equivalently to the dispersion of a tracer) in a single fracture. 
Several mechanisms contribute to the spreading of the front (or of the tracer). These include molecular diffusion, Taylor dispersion related to the velocity profile in the fracture gap and geometrical dispersion due to velocity variations between flow lines in the mean fracture plane  (see \citet{Drazer02} and \citet{Drazer04} for a discussion of the contribution of the above mentioned mechanisms to tracer dispersion).

Here,  we shall take only into account  the last mechanism : this allows us to assume  a two-dimensional flow field $v(x,y)$ obtained by averaging the three-dimensional velocities over the gap of the fracture with $v(x,y)=\langle v(x,y,z)\rangle_z$. The pressure gradient inducing the flow and the resulting global mean velocity are parallel  to the axis $x$ and the front (or the tracer) is initially located at the inlet of the fracture on a line parallel to the axis $y$. In the following, we analyze the development of the front (or the tracer line) with time, assuming that  its various points move at the local flow velocity, and we present a simple analytical model accounting for its geometry.  

This model is based on the results  discussed above, namely that the aperture field is structured into channels perpendicular to the shear displacement ${\vec u}$. For  a mean flow parallel  to these   channels, the tracer particles  follow paths that are only weakly tortuous; also,  the variations  of their velocity along  these paths are small compared to the velocity contrasts between the different channels.
Under  these assumptions,  the velocity of a particle located at a distance $y$ 
measured perpendicular to the mean velocity satisfies  : 
$\vec{v}(x,y) \approx v(y) \vec{n}_x$, where $\vec{n}_x$ is a unit vector parallel to the mean flow. Note also that there are no contact points between the walls of the fractures we have used : this avoids to take into account  the large tortuosity of the flow lines in their vicinity. For each channel  the velocity $\vec{v}(y)$ is related to the pressure gradient $dP/dx$ by Darcy's equation : 
\begin{equation}
\vec{v}(y)=-\frac{a^2(y)}{12 \mu} \frac{dP}{dx} \vec{n}_x,
\label{eq:velocity}
\end{equation}
 in which $a(y)$ is the equivalent (or hydraulic) aperture of the channel. 
\begin{figure}
\includegraphics*[width=\W]{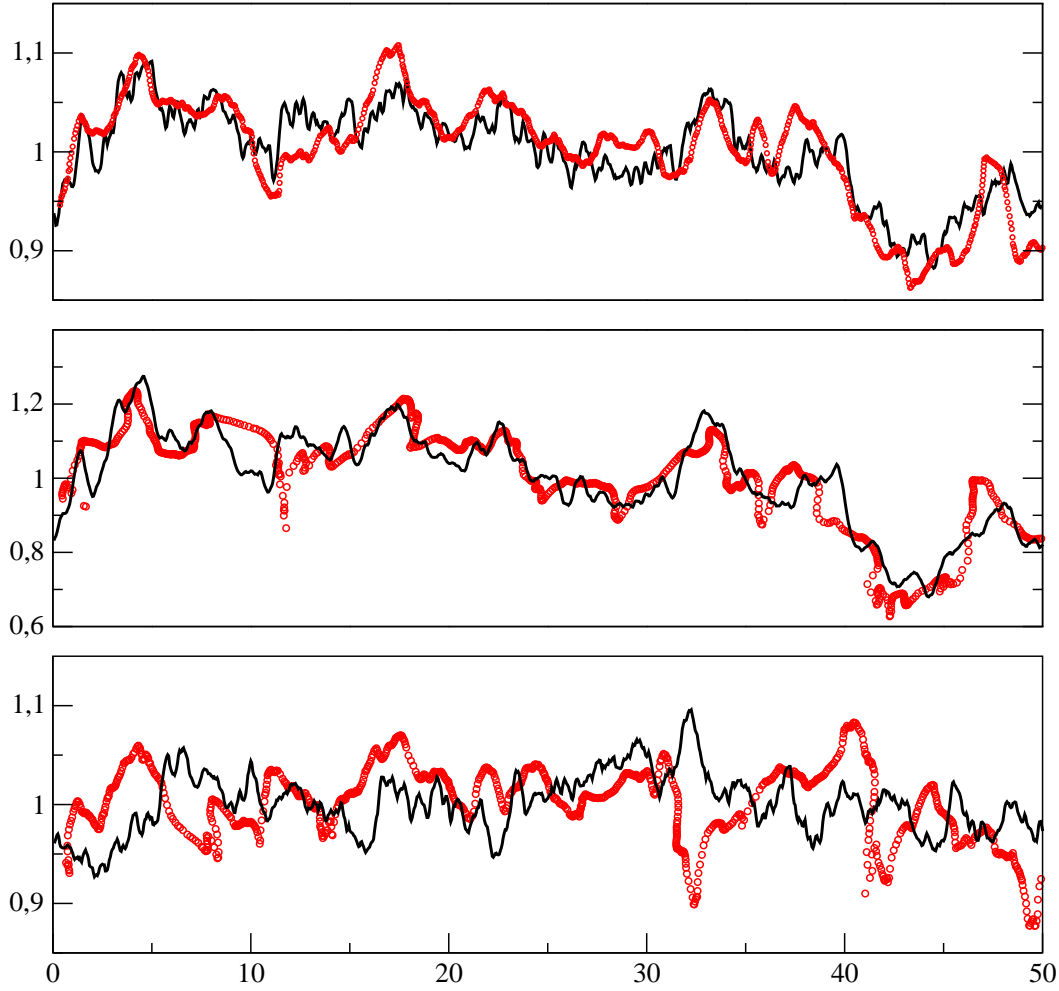}
\caption{Compared normalized front profiles  $\frac{x(y,t)}{\bar{x}(t)}$ (circles) and normalized apertures  $\frac{a(y)^2}{\langle a^2(y) \rangle_y}$ (solid lines) as a function of the distance $y$. across the flow. Top and middle graphs : flow normal to shear displacement $\vec u$ with  $u = 0.2\ mm$ and $u = 0.6\ mm$. Bottom graph :  Flow parallel to $\vec u$ for $u = 0.6\ mm$ (the $y$ scale of the solid line has been amplified by a factor $10$). For all graphs,  the fracture aperture is $a_0 = 1\ mm$ and its size is $51\ mm \times 51\ mm$.}
\label{fig:fronts}
\end{figure}
Then, particles starting at different distances $y$ at the inlet
 move at different velocities, $v(y)$, and their distance $x$ from the inlet at time $t$ after the injection satisfies $x(t)=v(y)\, t$;  the mean front position  moves then at 
the average velocity with $\bar{x}(t) = \langle v(y) \rangle_y \, t$.
Combining the previous relation with Eq. (\ref{eq:velocity}) we obtain:
\begin{equation}
\frac{x(y,t)}{\bar{x}(t)} = \frac{a^2(y)}{\langle a^2(y) \rangle_y}.
\label{eq:model}
\end{equation}  
Finally, previous studies have shown that, for relatively small aperture fluctuations, the hydraulic aperture is well
approximated by the geometrical aperture \citep{Zimmerman91,Brown87} : this suggests that $a(y)$ can be approximated by the average of the local apertures along the direction  $x$ with $a(y)=\langle a(x,y)\rangle_x$.

The validity of this  assumption has been tested by comparing  the shape of the front $x(y,t)$ determined from simulations using the lattice Boltzmann method to the
variations of the square $a^2(y)$ of the effective aperture. For that purpose, the  normalized displacements $x(y,t)/\bar{x}(t)$ at a time $t$ parallel to the mean flow of particles released initially at $x =0$ are plotted in Fig.~\ref{fig:fronts} as a function of the distance $y$ together with the normalized square of the aperture averaged parallel to the flow : $a^2(y)/\langle a^2(y) \rangle_y$.
When the mean  flow is perpendicular to the shear displacement $\vec u$, the variations with $y$ of the  normalized position of the particles and of the normalized square aperture are very similar both for $u = 0.2 mm$ and $u = 0.6 mm$ (top and middle curves). The small differences may be due in part to the effect of viscous entrainment between adjacent flow channels due to viscous diffusion. On the contrary, when the mean flow is parallel to $\vec u$ (bottom curves), there is very little correlation between the two curves and the amplitudes of the variations are very different (note that the curve corresponding to $a^2(y)/\langle a^2(y) \rangle_y$ has been magnified by a factor of $10$).

These strong differences reflect and confirm the anisotropy of the aperture field induced by the shear displacement $\vec u$ : the assumptions underlying the above model and, particularly, Eq.~\ref{eq:model} are indeed only valid if the flow field can be described as parallel channels with a length of the order of that of the fracture. For a  mean flow perpendicular to $\vec u$,  this model represents a good approximation :  the aperture of the flow channels displays long range correlations and particles are convected along 
each channel with a weakly fluctuating velocity.  Note that, in this case, the global width of the front parallel to $x$ increases linearly with time while, for dispersive processes it increases from the square root  of time. This difference results from the fact that the correlation length of the velocity of fluid particles is of the order of the total path length through the fracture; a dispersive regime can only be reached if the correlation length is much smaller.
For flow parallel to $\vec u$, the flow lines are more tortuous and velocity fluctuations are enhanced : the model does not account any more for the geometry of the front and its width parallel to the mean flow is in addition reduced.

\begin{figure}
\includegraphics*[width=\W]{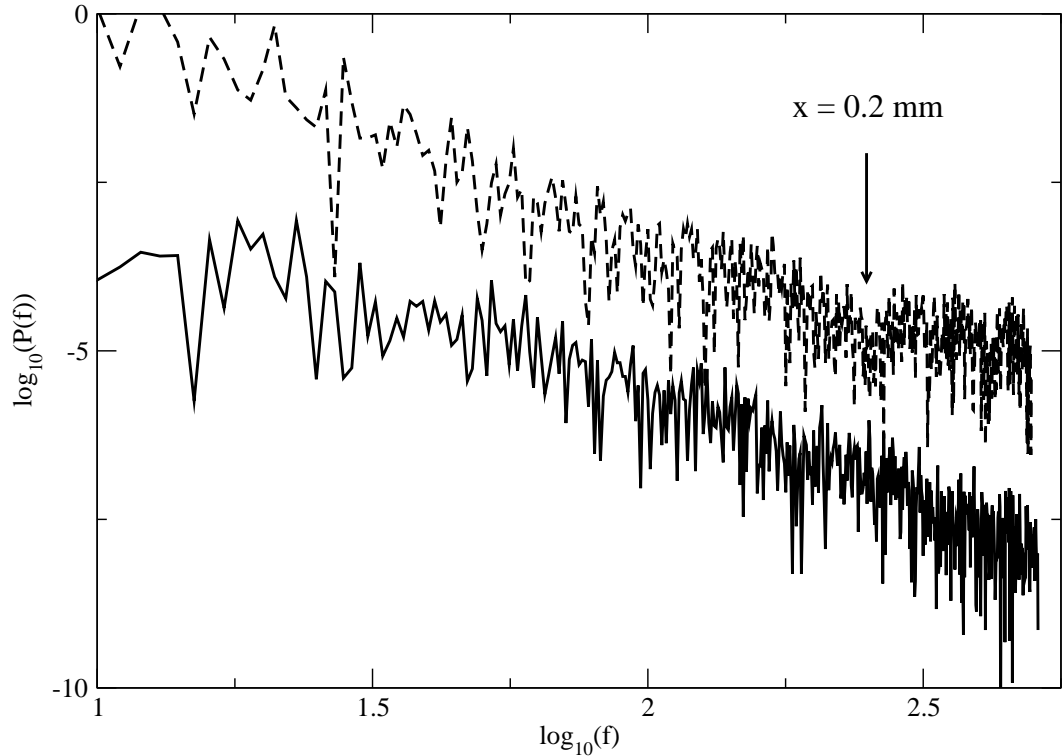}
\caption{Log-log plot of the power spectra  $P(f)$ of the deviations of the front displacement $x(y,t)$ (dashed line) and of the squared effective aperture $a^2(y)$ (continuous line) as a function of the spatial frequency $f=L/x$ (shear displacement $u = 0.2 \ mm$, mean aperture $a_0 = 1\ mm$, fracture size $L=51.2\ mm$). The curves are shifted vertically relative to each other for convenience.}
\label{fig:fourier}
\end{figure}

Further informations are obtained by performing  Fourier transforms of the deviations of both the displacement $x(y,t)$ of the front and the square $a^2(y)$ of the estimated effective aperture from their mean values (the mean flow is perpendicular to the shear displacement $\vec u$). The resulting power spectra are displayed in Fig.~\ref{fig:fourier} and, 
as could be expected from the qualitative analysis of the curves of Fig.~\ref{fig:fronts}, both spectra are similar up to frequencies $f$ of the order of  $f=L/u$. In fact, for wavelengths smaller that the shear displacement $u$, only  $a^2(y)$ displays  self-affine properties.
\section{Discussion and Conclusions}
The experiments and numerical simulations  realized  in the framework of the present program have revealed new features of flow and transport in rough fractures which may be of importance in the modelization of fractured field reservoirs, such as those encountered at the Soultz site. 
We have specifically studied the influence of a relative shear displacement of complementary wall surfaces on the aperture field as well as on flow and transport  between them. 
This is particularly relevant to geothermal reservoirs where fluid injections frequently induce relative displacements of the fracture walls. In agreement with many experimental results, the rough fracture walls have been assumed to display a self-affine geometry.

First, we have shown  that the relative shear displacements $\vec u$ induce an anisotropy of the flow field which reduces permeability for flow parallel to $\vec u$ and enhances it in the perpendicular direction; this may account in part for the permeability enhancement in some sheared zones of the reservoirs. These variations are well predicted by assuming the appearance in the aperture field of large ridges perpendicular to the shear and extending over the full length of the fracture : these ridges act as parallel channels enhancing flow parallel to them and reducing transverse velocity fluctuations. This anisotropy has been confirmed quantitatively by computing semivariograms of the aperture fields : these display a correlation length proportional to the shear displacement $\vec u$ for a given orientation  but significantly larger perpendicular to $\vec u$ than parallel to it.

A second important result is the strong correlation  between the roughness of the 
fracture walls and the geometry of the displacement fronts, particularly for  flow  perpendicular to $\vec u$. 
In particular,  the displacement front has a self-affine geometry characterized by  
the same exponent as that characterizing the  fracture walls. 
Also, for a mean flow perpendicular to $\vec u$, the front width is found to increase linearly with  distance : this reflects a channelization  parallel  the mean flow, resultinb in weak velocity fluctuations along the paths of the fluid particles. 
Another important consequence is the fact that the front geometry can be estimated accurately in this case  from the variations of the square of the aperture averaged parallel to the mean flow.

These results raise a number of questions that will have to be considered in  future studies. 
First, one may expect the
spatial correlations of the velocity field to decay eventually, leading to a normal Gaussian dispersion process at long distances.  This has not been observed in our experimental or numerical systems and we have currently no  estimate of the corresponding
correlation length. Also, the channeling effect  might result in a motion  of some fluid particles at a velocity  substantially faster than the average flow velocity, at least in transient regimes of duration shorter than the time necessary to sample the flow variations inside the fracture. This effect may have a strong influence on the thermal exchange between the rock and the fluid, or on  the transport of species in dissolution-deposition processes encountered, for instance, on the Soultz site. In the case of  a channelized flow, for instance, the variability of the residence times in the various flow channels may result in temperature inhomogeneities in the outflowing fluid.

\section{Acknowledgments}
We are indebted to G. Chauvin and R. Pidoux  for their assistance in the realization of the experimental
set-up.  Computer resources were provided by the National Energy Research
Scientific Computing Center.
HA and JPH are supported by the CNRS and ANDRA through 
 the EHDRA (European Hot Dry Rock Association)  and PNRH programs.
GD and JK were supported by the Geosciences Program of the Office of Basic
Energy Sciences (US Department of Energy), and by a PSC-CUNY grant.
This work was facilitated by a CNRS-NSF Collaborative Research Grant and by the PICS CNRS n¡ 2178.

\end{document}